# "The Gravitational Red-Shift"

R. F. Evans and J. Dunning-Davies,
Department of Physics,
University of Hull,
Hull HU6 7RX,
England.

J.Dunning-Davies@hull.ac.uk

## Abstract.

Attention is drawn to the fact that the well-known expression for the red-shift of spectral lines due to a gravitational field may be derived with no recourse to the theory of general relativity. This raises grave doubts over the inclusion of the measurement of this gravitational red-shift in the list of *crucial* tests of the theory of general relativity.

## Introduction.

In most of the standard texts concerned with the General Theory of Relativity [1,2,3], reference is made to three crucial tests of the theory; - the well-known Advance of the Perihelion of the planet Mercury, the Gravitational Deflection of Light Rays, and the Gravitational Shift of Spectral Lines. Weinberg [3] does draw attention to two other possible tests, but still retains the above three in his list.

The situation seems quite interesting as far as the third of these apparently *crucial* tests is concerned. In the first of the above references, attention seems to be concentrated solely on the effect being due to general relativistic effects. In the third reference, it is noted, quite clearly, that the effect arises due to the principle of equivalence alone; the Einstein Field Equations are not concerned in the derivation of the formula at all. It is only in the second reference listed here that attention is drawn to the fact that, as well as the test only being one of the validity of the principle of equivalence, alternative derivations of the result exist also. It is possibly not surprising, therefore, that few people seem to realise that the said result may be derived with no recourse to the general theory of relativity whatsoever, nor to the principle of equivalence. It is possibly surprising, therefore, to find that one of the alternative derivations mentioned in the second reference is this one in which the general theory of relativity plays no part. The end result is that the existence of this derivation is not widely known and leads to situations, such as that reported here, in which the result is derived afresh with no knowledge of the earlier work.

## The Gravitational Red-Shift.

Conservation of energy yields the fact that the sum of the kinetic and potential energies is a constant. If a particle of mass $m$ is moving under the influence of a gravitational field generated by a massive central body of mass $M$, Newton's law of gravitation shows that the potential energy is given by $-GMm/r$, where $G$ is Newton's universal constant of gravitation and $r$ is the distance of the particle from the central massive body.

However, what of the kinetic energy which, for a normal material particle, is taken to be one half the product of the particle's mass with the square of its velocity? Obviously, such a formula would not apply in the case of a 'particle' of light, which has zero mass. However, the kinetic energy of a photon is given by $h\nu$, where $h$ is Planck's constant and $\nu$ is the frequency of the photon. If the mass-energy relation

$$E = mc^2,$$

which relates the kinetic energy to the product of mass and the square of the speed of light, is introduced, then an 'effective mass' for the photon may be deduced and is given by

$$m = h\nu/c^2.$$

The equation expressing conservation of energy then becomes

$$h\nu - GMm/r = h\nu - GMh\nu/rc^2 = \text{constant}.$$

This equation immediately allows the well established expression for the gravitational red-shift to be deduced. For example, if as $r \to \infty$, $\nu \to \nu_\infty$, the equation of conservation of energy becomes

$$h\nu - GMh\nu/rc^2 = h\nu_\infty$$

or

$$\frac{\nu_\infty - \nu}{\nu} = -\frac{GM}{rc^2},$$

which is the desired result.

**Discussion.**

In the above derivation of the expression for the gravitational red-shift, no appeal has been made to any aspect of the theory of general relativity, not even the principle of equivalence. Hence, the question must be raised as to how and why the measurement of the gravitational red-shift could ever be considered a real test of general relativity, let alone a *crucial* test as is so often claimed? It seems surprising that, once the above deduction of an 'effective mass' for the photon was recognised, this simple derivation of the red-shift formula did not become more widely recognised. It is possibly even more surprising to note that, although the writing-up is slightly different, this derivation is included in the second reference listed, but, even there, it is still regarded as being a crucial test of general relativity. In truth, it would seem that the result for the red-shift of spectral lines due to the action of a gravitational field has nothing specifically to do with the theory of general relativity. It is a result which draws on more modern results due to such as Planck and Poincaré, but, apart from those, is deduced from notions of Newtonian mechanics alone. As such, it seems to have no place in a list of *crucial* tests of general relativity, although the theory of general relativity obviously must not contradict this result. This point has been made also by Lavenda[4] in an article where he also shows that, of the three usually listed crucial tests, the deflection of light and the perihelion advance can be treated as diffraction phenomena on the basis of Fermat's principle and the modification of the phase of a Bessel function in the short-wavelength limit. Incidentally, Lavenda also deals with the problem of the time delay in radar echoes by this means. Where does all this leave the General Theory of Relativity?

**References.**